\documentclass[final,abstracton]{article}
 
\usepackage{arxiv}

\usepackage[english]{babel}

\usepackage{graphicx}
\usepackage{wrapfig}

\usepackage{etoolbox}%
\usepackage{microtype}

\usepackage{siunitx}

\usepackage{csquotes}
\MakeOuterQuote{"}

\usepackage[unicode=true,breaklinks=true,colorlinks=false,linkcolor=black,pdfborder={0 0 0},pdfdisplaydoctitle=true,pdftitle={A holistic approach for rapid development of IIoT systems},pdfauthor={Tilman Klaeger, Konstantin Merker}]{hyperref}
\hypersetup{final}

\usepackage{cite}

\title{FastIoT -- A framework and holistic approach\\ for rapid development of IIoT systems}
\author{Tilman Klaeger, Konstantin Merker\\
	Fraunhofer Institute for Process Engineering and Packaging (IVV)\\
	Division Processing Technology\\
	Dresden, Germany \\
	{tilman.klaeger@ivv-dd.fraunhofer.de}
}

\begin{document}

\maketitle

\begin{abstract}
While lots of research has been conducted on the architecture of Industrial Internet of Things (IIoT) systems, concepts of structuring their development processes are missing. Therefore, we propose a holistic approach supporting organizations in rapid development of IIoT systems. It includes the structuring of the development process into multiple projects sharing project conventions. Utilizing a single configurable build script for all projects, our goal is to make the integration of code from various projects into broader IIoT systems easy and the systems decomposable with minimal effort. 
\end{abstract}

\section{Introduction}

Usually, the development of information and industrial automation systems have very different characteristics in terms of abstraction, reusability, (automated) testing and deployment.
One reason is the longer lifetime of hardware in industrial equipment compared to IT systems. Changes in automation structures are complicated because of regulatory aspects and safety checks. Looking at the field of industrial automation, IIoT systems are more and more spreading out and bringing new methods of software development closer to the hardware \cite{Liao2018IndustrialInternetThings}.
This opens up opportunities for faster and more agile development methods.
Also integration of latest software technologies like machine learning based applications is eased using IIoT architectures.
This will not only help with latest machines but also with older ones in the brown-field, where data science is still a major technological challenge \cite{Klaeger2021DataScienceIndustrial}.

On the software development side there are many possible ways how to structure software systems in general. During the last years, trends to compose systems based on small units, mostly known as microservices, have gained much interest \cite{DiFrancesco2018MigratingMicroserviceArchitectures}. As hardware becomes more efficient these technologies and aspects of development are pushing into new domains like IIoT. This creates opportunities but also challenges and things to consider for the development of such systems.

Some advantages of the new software architecture are more decoupled and better scalable systems. Following this trend, lots of new technologies have been established for containerized applications and for development operations and build automation, mostly referred to as DevOps \cite{FerreiraLeite2019SurveyDevOpsConcepts}. This all leads to a new approach of developing industrial applications and is eagerly needed for integration of so called "smart products" often integrating some sort of machine learning applications.

The reusability of microservices makes it possible to develop such systems split across multiple projects and even organizations. Individual parts of the system become decomposable and integrable in multiple applications.

However, despite these possibilities we found it hard to find a structured approach for the distributed development of systems in a practical way, where a service is developed once and might be used within different projects.

So, our focus is to provide a framework to help setting up projects and to allow a better decomposability of individual services.
During the lifetime of a software system, some functionality might be moved between projects. Also, some projects are discontinued, but a portion of the functionality is still maintained. Being able to transfer them to active projects and dismissing the not needed parts results in a low maintenance effort. To do so, it must be possible to develop individual services in different projects and being able to transfer them without causing large refactorings. The same applies to libraries, tests, configurations and documentations.

In this paper we show, how it is possible to develop IIoT systems split in multiple projects using our approach and combining them in multiple, distinct IIoT systems while maintaining the possibility to develop new code rapidly and still be very flexible.

The resulting open source framework "FastIoT" will thus lead a quicker project setup and fewer lines of code to get the actual task done.
Therefore concentrating on the development of the actual application logic within an IIoT project becomes much easier.

\section{Related work}

The related work can be categorized into general software development trends and proposed architectures for IIoT. Within the first category, current software trends like structuring software systems into small deployable units, also called microservices happen. There are approaches for migrating monoliths to microservices \cite{Lauretis2019MonolithicArchitectureMicroservices}. Microservices are shown to be useful for IoT with some limitations as well \cite{Lu2017SecureMicroserviceFramework}. They have been applied and show advantages in flexible reusability, on-demand scalability and to speed up the creation of remote sensing products \cite{Xiang2018UsingMicroservicesRapid}. In other domains microservices have shown potential as well, for example in health applications \cite{daSilva2019MicroserviceBasedApproachIncreasing}. Other research looks into the interoperability of microservice-based systems by proposing a microservice-based reference architecture and a reference implementation \cite{Yuan2019ArchitectureInteroperabilityRepeatability}.

Another trend related to microservices is the availability of Open-Source technologies usable for microservice architectures. There is research conducted combining them in a framework \cite{Akasiadis2019MultiProtocolIoTPlatform}. Deploying microservices using the container runtime Docker for IIoT-Applications with resource-constrained devices has shown to be possible  \cite{Jaramillo2016Leveragingmicroservicesarchitecture, Rufino2017Orchestrationcontainerizedmicroservices}.

Within the second category, there are framework proposals for IIoT.
Some of them are focusing on special aspects like protecting personalized privacy data for mobile crowdsensing. It includes a rational data uploading strategy where users can choose their preferred privacy level \cite{Xiong2020PersonalizedPrivacyProtection}. Another framework focus on the integration of several subsystems via an intelligent API layer \cite{Uviase2018IoTArchitecturalFramework}. Other authors focus on a  framework utilizing blockchain technology to provide protection against data confidentiality attacks \cite{Medhane2020BlockchainEnabledDistributedSecurity}.
More general aspects focus on distribution of the service \cite{Kurte2020DistributedServiceFramework} or the scalability and security \cite{Amoretti2021ScalableSecurePublish}.
Panayiotou et al. present a library making integration with message brokers within an IIoT context easier and focus on protocol abstraction but not on development of whole microservice architectures \cite{Panayiotou2022Commlibeasytousecommunication}.
Sun et al. propose an open IoT framework based on microservices without going into developing details but focus on the overall architecture \cite{Sun2017openIoTframework}.

The list may be continued, but all these frameworks focus on the software architectures, but not on the development aspects of these systems.
We therefore consider an easier stage up process mandatory.

\section{Challenges to be solved}

While service-oriented systems have advantages in terms of scalability and composibility, major challenges result in the development of such systems.
Whereas monolithic applications may include functionality via external libraries and can be developed in a single repository, microservices are typically developed in multiple projects and the deployment spans across multiple nodes, heavily relying on external services like message brokers, databases, monitoring services and further tooling.

The creation of new projects including microservices, prototypes and test benches are a common task in many software departments developing microservices.
A quick way to setup projects, configuring services and managing of deployments needs to be provided by a framework to be successful.
Also refactoring processes may involve moving libraries and single services between projects. This needs to be considered implementing the framework.

Deploying software in an IIoT context can be done in different ways.
Whereas smallest architectures relay on micro controllers like the well known ESP32 by Espressif Inc., many projects use bigger hardware based on ARM or even x86 architectures often running Linux as operating system.
The proposed framework focuses on the later and thus can utilize containers for deploying the services.
Building containers for distinct architectures is tedious but should be possible to be covered by a well designed framework and its tooling.

Various other tasks may occur in specific projects.
The list may include converting API specifications, creating specific export files or configurations for a CI runner.

Interchanging data between services is essential for a service-oriented architecture and therefore needs special attention.
An easy way of data modeling is also needed to help structuring data passed between the services within a project and between services of different projects.
Especially for less skilled software developers forcing the step of data modeling helps to structure thinking and project design.
Closely related is the overall message handling and data serialization to pass messages between services and the central message broker.
For rapid development there should be no need to think about serialization mechanisms as it should be abstracted by the framework.

\section{Approach}
The tasks to be solved mainly divide into two parts: Controlling project setup, deployment management and building the containers for the microservices utilizing the CLI on the one side.
On the other side the common library to help with data modeling, message serialization and transmission but also supporting in common tasks like database connectivity.

\subsection{Common project structure}

To make transferring of functionality between projects possible, each project must be capable of handling a broad range of features and stay within a given and mostly fixed structure.
This includes options for a custom library, to be used and integrated in the project services but potentially even in other projects.
A project also may contain multiple services, automated tests, configuration files for deployment, documentation and some other project-related files.
It is possible to have projects only containing a small range of the named features for example only a library, but they are in the same way structured as projects also including more features like services.

\begin{figure}[h]
	\centering{\includegraphics[scale=0.6]{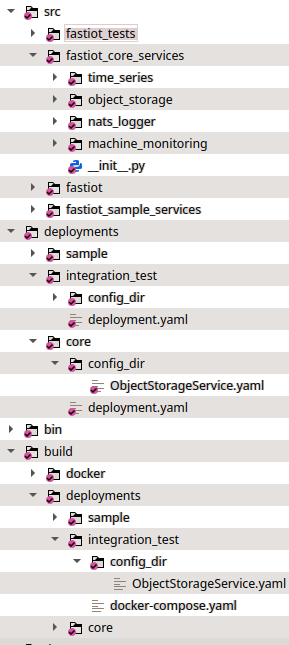}}
	\caption{Project related files of SAM-Project in a single source code repository.}
	\label{fig1}
\end{figure}

Different use cases also require different features, but for the IIoT domain in the context of machine learning, this set of features covers most requirements. Furthermore, it is possible to add features for projects in a similar way in the future, if required.
The framework itself uses the proposed structure, so working on a project based on the framework and developing the framework itself falls under the same principals.
This common structure helps developers when switching between projects but also an easier handling from the CLI and from some kind of build automation system.
The project structure is shown in Fig.~\ref{fig1}.

The figure shows, where certain files are typically found within the project.
Deployment configurations, FastIoT library, services, tests, beside some other files are located following a convention.

Working with sensible defaults or "convention over configuration" will help keeping the configuration small, but it will not completely make it obsolete. A good set of those sensible defaults therefore has to be implemented. In our understanding, a sensible default will work in the majority of cases, but will also respect the open-closed principle, therefore the default configuration is extendible with custom functionality where needed.
This is implemented using a plugin architecture utilizing Python's subclasses feature.

An extendable command line interface (CLI) is provided for common tasks e.g. new project setup, building of containers and deploying configurations.

\subsection{Common data models, serialization and message passing}

To interchange data between services of different projects the data models, especially for standard cases like sensory data needs to be fixated.
Adding new data types to own projects can easily be done using Python objects based on the popular library "Pydantic".
Sharing data types across projects can be done by providing a library to be integrated into other projects working with same data types.
The framework itself provides basic data types for "Things" (sensor or actuators), more specific data types therefore need to be created in the individual projects.

Even though the framework itself is not designed with hard-realtime requirements in mind the overall performance for faster running processes is a relevant criterion.

To transfer data between services the data needs to be serialized as it is not easily possible to transfer Python objects.
We don't recommend using Python's method "pickle" with the ability to transfer instantiated classes \cite{Tangmunchittham2022AnalysisPythonSerialization}.
Sending memory-dump alike Python objects result in very short serialization time but can be incompatible between different Python versions (and CPU architectures). It also removes the possibility to develop services in different programming languages as in other programming languages the parsing of Python binary data is for practical purposes non-existent.
Common for webservices is JSON ("Java Script Object Notation") but being relatively slow compared to msgpack \cite{Gentz2019SPARCSStreamProcessingArchitecture}. We therefore propose JSON for webservices e.g. a web-based dashboard in the context of IIoT and msgpack for data transfer between services. The framework provides both serialization possibilities. Building on Pydantic it is also possible to extend custom serialization methods.

Comparing different message brokers "nats.io" can be considered one of the fastest message brokers and was therefore chosen as a first class citizen \cite{Suri2019ExperimentalEvaluationGroup}.

\subsection{Project setup and management using the CLI}

Each project starts with a setup phase.
The project structure and common source code and configuration files can be generated based on templates.
The generation process is conducted via a CLI and can be customized to match the project needs and provide a good starting point for development, optionally including sample services with method stubs.

\subsection{Development of microservices}

After setting up the project layout, application logic needs to get implemented by the developer.
In the context of IIoT, domain logic this is often not integrated by highly skilled programmers, but instead by experts from different domains like data scientists or automation engineers.
Thus, complex topics need be encapsulated and abstracted to ease the overall development process.
Those topics include but are not limited to microservice handling, data models and data serialization for messaging between the services. Functionality for managing the connection to various services like databases, message brokers etc. are supported by shared libraries within the framework.
Handling configuration files for specific services like hardware configuration as well as framework-wide environment variables can be imported using framework libraries as well.

Each service can indicate deployment requirements. These can be ports, devices, databases, etc. and restart policies. For easier fault-recovery containers are typically configured with an always-restart policy. The framework also includes exception handling functionality as in modern programming languages like Python, many tasks run asynchronously and the built-in exception handling might not redirect exceptions properly and instead drop them. This is often the case with Python’s async library. Thus, the framework handles exceptions in a way to cause the program to restart. This is particularly the case of any issues with connectivity to hardware, services or other unexpected behavior.
Overall this will result in robust services simply restarting on any faults if those are not caught by internal logic.

It is also possible to generate services based on templates or construct services inheriting from an abstract base class implementing common functionality. Doing so speeds up the development.

\subsection{Testing}

A major task during development is testing. Automated tests like unit tests or integration tests play an important role for good quality software and therefore preparations are made for easy setup of tests.
Specific test configurations allow to start containers with broker and other services like databases for local automated testing and automated testing on a Continuous Integration Service (CI-runner) like Jenkins.
A testing library is included to read the broker and other connection parameters with their usernames, password and especially local ports.

Writing automated tests seems not to be necessary for small projects or prototypes with a foreseen lifetime.
Instead, having fast results and iterating on them is more important. To do so, starting services locally can be used for debugging while being able to connect to a local broker in a container or a central broker on a development server. This can be done by setting environment variables. They can be set either directly in the integrated development environment (IDE) or via importing a specific "env-file" in the IDE containing all environment variables.
This also helps people working more in a trial-and-error type of developing.

To start the test dependencies the CLI may be used just as for running the actual tests.
Starting the tests from a IDE is possible as well, as the testing library will read the configuration when the tests are started.

\subsection{Building services as containers}

Especially in the IIoT domain different processor architectures need to be targeted.
Using plain Python code and a runtime compiled and running on the selected target is one option.
As the framework makes extensive use of containerized applications this does not work but instead containers for all target platforms need to be built.
Using the "buildx" command provided by Docker and the emulator "qemu" it is possible to build cross-platform containers.
At the moment building and exporting to the local machine allows only for one architecture, when pushing to a registry multi-platform labels can be built.
The whole process of setting up qemu and docker is covered by the CLI, so only the service needs to be configured to run on different architectures.

\subsection{Deployment configurations and semi-automatic deployment}

The focus is on reusability of service across projects.
Therefore, each project is be able to use other projects services.
To do so, each project has one or many deployment configurations.
The configuration is in a specified format including services and custom environment setups for the services, if needed.
It is also possible to include service many times with differing configurations and environment variables.
Project-internal services can be built using the CLI whereas external ones are just included with their corresponding project and image name in the deployment.
As a project may contain many different deployments, it is possible to setup, for example, a server with some service and single board computers with distinct services.
Differentiating between development/ debugging setups and production can be achieved by using different image tags for debugging and production respectively.
Optionally, configurations may include hosts it should be deployed on.

Using the CLI the deployment configurations will be translated to a docker-compose file, adjusted files with environment variables (".env-files") and other specific configuration.
This directory can easily be copied to the target devices.
To automate this task the orchestration tool "Ansible" can be utilized and is therefore integrated within the framework.
With a single command a final deployment configuration is copied to all specified hosts and executed there.
This makes setting up a bunch of similar configured devices very easy.

\section{Concept Evaluation}

\subsection{Setup Time}

For an experienced developer the whole setup process from zero to a project can be done in 10 minutes starting from a fresh virtual machine. This includes creation of the project, creating a central repository for the source code, registering the project in the CI-runner and deploying two microservices communicating with each other locally.

A local setup can boil down to seven commands including the setup of a new virtual environment for Python to build a containerized service sending and receiving data within a small IIoT deployment.

\subsection{Lines of Code for idiomatic basic consumer and producer services}

Another focus is on less overhead an quick development of new services with a focus on the actual application logic.
Therefore fewer lines of code for an individual service are in general favorable.
To count the lines of the code for different parts of a freshly created project the Python-tool "pygount" was used.
Using the CLI a service can be created with code for sending and receiving some example data together with the corresponding service configuration.
Also included in this code is some logging output and some comfort functions.
This code can be stripped down to only sending (producer) or receiving (consumer) data as a minimum viable service.
The resulting lines of code can be seen in table \ref{Tab:LinesOfCode}.

\begin{table}[h]
	\caption{Lines of code for different parts of a project with plain Python code and full code including configuration files.}
	\label{Tab:LinesOfCode}
	\begin{tabular}{lrr}
		\textbf{Project Part} & \textbf{Lines of Python Code} & \textbf{Total Lines of Code} \\
		\hline
		Freshly created project w/o services & 11 & 25 \\
		\hline
		Service created by template & 34 & 38 \\
		\hline
		Stripped down consumer & 9 & 13 \\
		\hline
		Stripped down producer & 18 & 22 \\
		\hline
	\end{tabular}
\end{table}

With only 9 lines of code, all them created by a template if used, a developer can quickly focus on the actual application logic without having to manage the typical microservice overhead like message passing.

\subsection{Messaging performance}

For a basic test of the performance two services where created to send and receive data of the format "Thing" containing basic sensor-like values with a name, machine name, value and timestamp. The message size after serialization was \SI{154}{bytes}.
The time delay between sending messages was set to \SI{10}{ms} thus resulting in about \SI{100}{messages/second}.
The performance was evaluated by comparing the sent timestamp on reception with the current time on 1000 repeated messages.
As all containers for sender, receiver and broker where running on the same machine there is no influence of time synchronization and no delay due to network transfers included.
Some small delays due the local Linux networking stack used may be included but should not take big portions and are not avoidable anyways.

The test was run an a data science workstation with two Intel Xeon Sliver 4114 CPUs (\SI{2.2}{GHz}) running Linux Kernel 5.19.8 and Docker in version 20.10.18. Another Test was run on an Raspberry Pi 4 with Kernel 5.4.0 and Docker in version 19.03.13.

On the Workstation the time measured was at \SI{1.8}{ms}.
On the Raspberry Pi it took \SI{3}{ms} from message creation till decoding.

The chosen message serialization and broker show an overall good performance.
If the components chosen are the best performing ones is hard to tell and may be subject to further research if more performance is needed.

\subsection{Other findings and considerations}

Creating the sample services do help to quickly start a project but are tedious to maintain. While it is good to look for backward compatibility when adding new functions or refactoring some parts of the code, this does not always work. Whereas older services can simply stick to an older core library, projects just generated from scratch will use the latest library. Refactoring parts of the code is easy with modern IDEs, but those will usually omit templates. If the templates are not adjusted to changes in the core library this will cause bugs in fresh services and especially novices in the code will have a hard time to find the bugs. A possible next step would be to also test the templates using automated testing techniques.

Also, we found the correct design of data types to be very important in this field. Especially data types used by many services like sensor data must be designed very carefully.
Changes in data types require at least rebuilding - if not refactoring of all services using this data type for communication.
An alternative are custom data types for specific projects, but this is unfavorable because it denies the ability to transfer services between projects easily. Multiple data types for the same entity also cause higher maintenance effort.

A minor difficulty lies within the dependency between services in a way that sending data often causes services to wait until another service can use the data. Otherwise it is lost. As an example, a service reads out the process image of a machine controller (PLC) and sends changed values further on. If no suitable listener is available when sending the initial process image, the data will be lost forever. 
This is an issue because some values don not change often.
The question is, what defines a listener to be suitable? Is a service just receiving the data suitable? Or a service storing the data? What, if in a special case, there is no need for storing the data in a deployment? This is still an open issue as there are many corner cases.
Using retention in the broker this could be covered but will slow down performance and needs further evaluation in actual use cases.

We also found, that having similar project setups, it's possible to use a shared build pipeline. So not each project needs its own build pipeline which needs to be maintained, but instead one pipeline can be shared and therefore the maintenance becomes easier.

\subsection{Conclusions}

We presented a holistic approach to manage the development of software systems across multiple projects in the IIoT domain. We showed how it supports the development, building and deployment of micro services across various, distinct systems and how functionality can be moved between projects.

The framework presented is based on ideas and findings from past projects where a similar framework was used.
Thus we employ the explained concepts for various projects in the domain of self-learning assistance systems for machine operators in the food industry \cite{Rahm2018KoMMDiaDialogueDrivenAssistance} or paper production \cite{Schroth2022Optimizationpaperproductiona}, for a protein database, adaptive cleaning solutions \cite{Yin2021Conceptselflearningadaptive} and beyond and conclude, that it works very well for setting up new projects and integrating results from various projects together. Similar project layouts allow researchers and developers to move from one project to the other quickly and it is possible to transfer functionality or services between projects easily. Typical tasks in IIoT like hardware interfacing with machine controls, still often being quite a hassle, can thus be reused and also help for a lot of brown-field applications as proposed in \cite{Klaeger2021DataScienceIndustrial}.

The presented approach reduces effort for developers and researchers to move between projects. Additionally, the usage of templates reduces the time needed for the setup of projects and services. The sharing of features like the build pipeline reduces the amount of work for single projects and decreases the maintenance effort. Shared libraries can evolve over the lifetime of multiple projects and thus reduce the cost for new projects further.

Even though much complexity of such systems is covered and moved to abstraction layers, learning to work with CLI and framework still takes some time for people new to the field.
Also for debugging some knowledge in docker an containers has proven to be useful but not absolutely necessary.
Documentation is helpful for the first steps but is not yet able to cover all complexity arising when using multiple setups with various machines, services, technologies. Understanding the infrastructure handling deployments is still not trivial and needs skills in various domains from programming to IT infrastructures.

\newpage

\bibliographystyle{myIEEEtran}
\bibliography{z_citations.bib}

%
%
%

\vfill

\begingroup
\setlength{\intextsep}{0pt}
\setlength{\columnsep}{3pt}
\parindent0pt

\begin{minipage}[l][1.2in]{\textwidth}
	\begin{wrapfigure}{l}{1.1in}
		\includegraphics[width=0.88in,height=1.1in,clip,keepaspectratio]{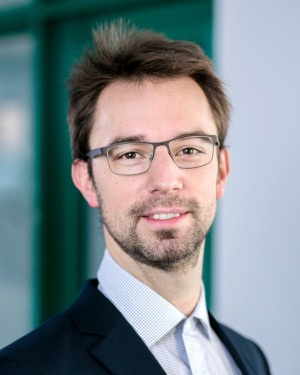}
	\end{wrapfigure}\par
	\textbf{Dipl.-Ing. Tilman Klaeger} studied mechatronics at the Technische Universit{\"a}t Dresden and started working at Fraunhofer IVV in 2016 and is now in the lead of the team data science. His major topic in research is machine learning on industrial data collected from packaging machines and processes.\par
\end{minipage}

\begin{minipage}[t][1.2in]{\textwidth}
	\begin{wrapfigure}{l}{1.1in}
		\includegraphics[width=0.88in,height=1.1in,clip,keepaspectratio]{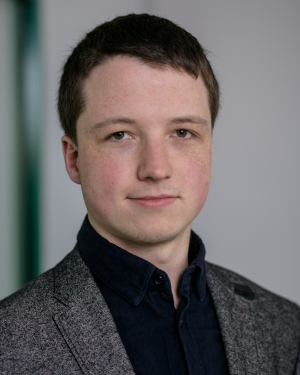}
	\end{wrapfigure}
	\textbf{Dipl.-Wirt.-Inf. (FH) Konstantin Merker} studied business informatics at the HTW Dresden. He is working in the team data science at Fraunhofer IVV since 2019. One focus of his research is on software architectures and development of (IIoT) applications.
\end{minipage}

\endgroup

\end{document}